\documentclass[final,
secnumarabic,nobalancelastpage,amsmath,amssymb,
twocolumn]{elsarticle}

\usepackage{lineno,hyperref}
\usepackage{graphicx}
\usepackage{bm}
\usepackage{float}
\usepackage{amsmath} 
\usepackage{amssymb}
\usepackage{xcolor}

\newcommand{\mL}{{\mathcal{L}}}
\newcommand{\mE}{\ensuremath{\mathcal{E}}}
\newcommand{\mB}{\ensuremath{\mathcal{B}}}
\newcommand{\lambdabar}{{\mkern0.75mu\mathchar '26\mkern -9.75mu\lambda}}
\newcommand{\req}[1]{Eq.\,(\ref{#1})} 
\newcommand{\rs}[1]{section~\ref{#1}} 
\newcommand{\rf}[1]{Fig.\,\ref{#1}} 
\newcommand*{\doi}[1]{\href{http://dx.doi.org/#1}{doi: #1}}

\journal{Physics Letters B}

\begin{document} 

\begin{frontmatter}

\title{Emergence of periodic in magnetic moment effective QED action}
\author{Stefan Evans\corref{mycorrespondingauthor}}
\cortext[mycorrespondingauthor]{Corresponding author}
\ead{evanss@email.arizona.edu}

\author{Johann Rafelski\corref{}}

\address{Department of Physics, The University of Arizona, Tucson, AZ 85721, USA}

\begin{abstract} 
We evaluate for the inhomogeneous static electric Sauter step potential the imaginary part of the emerging homogeneous in electric field effective Euler-Heisenberg-Schwinger action sourced by vacuum fluctuations of a charged particle with magnetic moment of arbitrary strength. The result is convergent for all values of gyromagnetic ratio $g$, periodic in $g$, with a cusp at $g=2$. We consider the relation to the QED beta-function which is also found to be periodic in $g$. We confirm presence of asymptotic freedom conditions using this novel method and document a wider range of $g$-values for which asymptotic freedom is present.
\end{abstract}

\begin{keyword}
Nonperturbative QED \sep Strong fields \sep Anomalous magnetic moment \sep 
Spontaneous particle production 
\\ \vspace{0.1cm}
arXiv:2203.13145
\end{keyword}

\end{frontmatter}

\section{Introduction}

For elementary charged leptons, an effective value of gyromagnetic ratio $g\ne 2$ arises on account of higher order Dirac $g=2$ particles interacting with the virtual radiation field. Moreover, should some of present day \lq elementary\rq\ fermions turn out to be composites of more elementary components, a value $g\ne 2$ could be in general expected. The anomalous magnetic moment is thus seen as an effective value. It is in this sense that we can say that there is no elementary particle which has exactly the Dirac gyromagnetic ratio $g=2$. 

We explore the effect $g\ne 2$ has on the quantum vacuum structure in strong fields, with the objective of generalizing the Euler-Heisenberg-Schwinger (EHS)~\cite{Heisenberg:1935qt, Weisskopf:1996bu, Schwinger:1951nm} effective action to $g\ne 2$ . In prior consideration, an effective vacuum response as a modification to the proper time formulation~\cite{Schwinger:1951nm} yielded a convergent expression in the domain of $|g|\le 2$~\cite{OConnell:1968spc, Dittrich:1977ee, Kruglov:2001dp, Kruglov:2001cx}: The proper time integral defining the final result diverges for $ |g| >2 $ after renormalization and cannot be regularized. Clearly this excludes the case of most if not all elementary fermions which due to vacuum fluctuations have an effective value $g>2$. 

For the case of magnetic-dominated EHS action with pseudoscalar $\mE\cdot\mB=0$, 
an alternate method applicable to $|g|>2$ consists of using the Weisskopf summation of Landau energy eigenvalues method~\cite{Weisskopf:1996bu}. The discrete spectrum is explicitly periodic in $g$, and yields a convergent result for $|g|>2$~\cite{Rafelski:2012ui}.  The action exhibits a cusp at $g=2$,  which appears in the beta function. Expecting the same behavior to occur for electric fields, in this work we obtain the imaginary part of the EHS type effective action for inhomogeneous electric fields generated by a static Sauter step (SS)~\cite{Sauter:1932gsa} potential. By considering the appropriate limit of the SS case we demonstrate the same periodicity in $g$ for the case of electric fields as was seen in the magnetic field case. Beyond the limit of quasi-constant electric fields, the Sauter inhomogeneous potential step allows us to explore the $g$-dependent vacuum response for the interesting case of localized sharply peaked electric fields.

In our approach we are generalizing Nikishov's imaginary  part of spin-$1/2$ action~\cite{Nikishov:1970br} to arbitrary values of gyromagnetic ratio $g$. We consider a second order relativistic wave equation, the so-called Klein-Gordon-Pauli (KGP) equation in the presence of the SS potential. KGP arises considering the product of the Dirac equation with its negative mass counterpart, allowing in this \lq\lq squared\rq\rq\ form for arbitrary magnetic moment. Schwinger employed the KGP form to obtain the proper time evolution operator in deriving EHS action, Eq.\,(2.33) in~\cite{Schwinger:1951nm}. Since then, the KGP equation was studied in the context of weak interactions by Feynman and Gell-Mann~\cite{Feynman:1958ty}, and in the context of anomalous spin $g$-factor by Veltman~\cite{Veltman:1997am}. 

Description of $g\ne 2$ by the KGP equation does not have an equivalent first order Dirac-like form. However, often in use is  another approach based on the Dirac equation: One can augment the Dirac equation by an anomalous magnetic moment term proportional to $g-2$. In the context of quantum physics, the differences between KGP and such first-order Dirac-Pauli (DP) effective formulation were explored in~\cite{Steinmetz:2018ryf}, which work is offering a comparison of the both methods. This study demonstrates the advantages and elegance of the KGP approach as compared to the incremental DP approach.

In the context of quantum vacuum structure study we choose the KGP form because the magnetic moment interaction is dimensionless as is prerequisite for a renormalizable quantum field theory~\cite{Veltman:1997am,Angeles-Martinez:2011wpn}, see also the more recent report~\cite{Espin:2015bja}. Making this choice we align our work with that of Kruglov's extension of the Schwinger proper time action operator~\cite{Kruglov:2001dp} to $g\neq2$. The KGP method is further supported by the limit $g\to 0$ which approaches the spin-0 behavior, up to a factor 2 related to the available two spin polarizations for spin-1/2 particles, and an overall sign due to quantum statistics.

With our results for the case of pure electric fields we obtain the $|g|>2$ periodicity as was predicted in the magnetic action study~\cite{Rafelski:2012ui, Evans:2018kor}. We note that therefore our result validates the postulated analytical continuation from the pure magnetic case to arbitrary constant electro-magnetic field cases offered in Ref.\,\cite{Rafelski:2012ui}. This is so since the effective action for constant arbitrary electromagnetic fields can only depend on eigenvalues of the EM-field tensor that are independent of the reference frame of the observer -- this argument was used in the first derivations of EHS effective action by Euler and Heisenberg~\cite{Heisenberg:1935qt},  Weisskopf~\cite{Weisskopf:1996bu}, and appears explicitly in the proper time method of Schwinger~\cite{Schwinger:1951nm}. 

We solve in~\rs{KGPSauter} the KGP equation in the presence of the SS potential. In~\rs{periodicE} we apply these results to computation of the imaginary part of effective SS action, {\it i.e.} vacuum persistence probability. We show that the periodicity in $g$ arises: i) taking the SS potential to quasi-constant electric field EHS limit, and ii) in the case of wide potential steps with heights barely above the $2m$-pair-producing threshold: In the latter case the EHS and Sauter field based actions differ by many orders of magnitude~\cite{Evans:2021coy}. 

In~\rs{betaCusp} we evaluate the charge renormalization governing beta-function using the strong field limit method of EHS effective action introduced by Ritus~\cite{Ritus:1975cf}, see also review work by Dunne in~\cite{Dunne:2004nc}. Our result extends the known $|g|\le 2$ result~\cite{Angeles-Martinez:2011wpn} to $|g|>2$. We demonstrate   the cusp at $g=2$, and the domains of $g$ far from the Dirac value $g=2$, in which the Abelian QED theory with $g\ne 2$ we call gQED exhibits asymptotic freedom. We discuss the physics context in which our results are important in~\rs{sConcl}, and consider theoretical and experimental implications of the here presented results and potential future  work.

\section{Solution of the KGP equation for the SS potential}
\label{KGPSauter}
 
The Sauter step potential has the explicit form
\begin{align}
A^\mu=&\;(V\;,\; \vec 0)
\;,\qquad
V=\frac{\Delta V}2\tanh[2z/L]
\;,
\end{align}
where $\Delta V$ is the step height and $L$ is the step width. The electric field points in the $z$-direction:
\begin{align}
\label{ESauter}
\mE=-\frac{\Delta V}L \mathrm{sech}^2[2z/L]
\;.
\end{align}
This field has a known solution for the Dirac ($g=2$) and Klein-Gordon ($g=0$) equations, which for pair-producing fields $e\Delta V>2m$, provides the imaginary part of effective action and particle production rate~\cite{Nikishov:1970br}. We follow the notation in~\cite{Chervyakov:2018nr} extending the derivation therein to $g\neq2$.

The KGP equation for the  wavefunction $\phi$ with a SS potential in the Weyl representation reads 
\begin{align}
\label{Cherv10}
&\;\Big[\partial_z^2-(m^2+p_\perp^2)+\Big(\omega-\frac{e\Delta V}2\tanh[2z/L]\Big)^2
\nonumber \\
&\quad
+ i(g/2)s \frac{e\Delta V/L}{\cosh^2[2z/L]}\Big]\phi=0
\;,
\end{align}
where $\omega$ is the incident particle energy, spin projection $s=\pm 1$, and gyromagnetic ratio $g$ can take arbitrary values. $\phi$ comprises scalar $\phi_0$, multiplied by normalized Weyl 2-spinor $X^{\mathrm{W}}$:
\begin{align}
\phi=\phi_0X^{\mathrm{W}}\;,\qquad
X^{\mathrm{W}}= 
\begin{pmatrix} 1 \\ 0 \end{pmatrix}\;,\;
\begin{pmatrix} 0 \\ 1 \end{pmatrix}
\;.
\end{align}
The asymptotic ($z=\pm\infty$) momentum states
\begin{align}
p_{z\pm}=\sqrt{(\omega \mp e\Delta V/2)^2-m^2-p_\perp^2}
\;.
\end{align}
Using notation
\begin{align}
\mu=&\;\frac{L}{4}p_{z-}\;,\quad
\nu=\frac{L}{4}p_{z+}\;,\quad
\lambda=\frac{L}{4}e\Delta V 
\;,
\end{align}
\req{Cherv10} becomes
\begin{align}
\label{Cherv22}
&\;\Big[\partial_z^2+\frac{4/L^2}{\cosh^2[2z/L]}\Big(
\nu^2(e^{4z/L}+1)+\mu^2(e^{-4z/L}+1)
\nonumber \\
&\;\qquad\qquad\qquad\qquad
-\lambda^2-i(g/2)s\lambda
\Big)\Big]\phi_0=0
\;.
\end{align}
We then apply\\[-1cm]
\begin{align}
\xi=&\;-e^{-4z/L}
\;,
\end{align}
and redefine the wavefunction as
\begin{align}
\label{redefW}
\phi_0=(-\xi)^{-i\nu}(1-\xi)^{i\chi}w
\;,
\end{align}
where $\chi$ is independent of $z$ and will be determined below. Using \req{redefW} we rewrite~\req{Cherv22} to give
\begin{align}
\label{lastWeyl}
&\;\xi 
\Big((1-\xi)w^{\prime\prime} -2i\chi w^\prime
+\frac{i\chi (i\chi -1)}{1-\xi}w
\Big)
\nonumber \\
&\;
+(1-2i\nu) \Big((1-\xi) w^\prime -i\chi w\Big)
\nonumber \\
&\;
+\Big( \frac{\lambda^2+i(g/2)s\lambda}{1-\xi} -(\mu^2-\nu^2) \Big)w
=0
\;.
\end{align}
To recast~\req{lastWeyl} into a hypergeometric differential equation, the divergence at $\xi=1$ must be removed following the procedure of~\cite{Chervyakov:2018nr}, which we modify to account for $g$. We require
\begin{align}
i\chi (i\chi -1)=-\lambda^2-i(g/2)s\lambda 
\;, 
\end{align}
solved by 
\begin{align}
\label{chiform}
\chi[s]=-i/2+s\sqrt{\lambda^2+i(g/2)s\lambda -1/4}
\;\underset{g=2}\longrightarrow \;
 s\lambda 
\;, 
\end{align}
which has the useful property 
\begin{align}
\label{conj}
\chi^*[s]=-\chi[-s]
\;.
\end{align} 
Plugging~\req{chiform} into~\req{lastWeyl}, the differential equation simplifies to 
\begin{align}
\label{lastlastWeyl}
&\;
\xi (1-\xi)w^{\prime\prime} +\Big((1-\xi) (1-2i\nu)-2i\xi\chi \Big)w^\prime 
\nonumber \\
&\;
+\Big((\chi -\nu)^2 -\mu^2\Big)w
=0
\;.
\end{align}
\req{lastlastWeyl} is solved by the hypergeometric function
\begin{align}
\label{wform}
w=F_{21}[\alpha,\beta\,;\,\gamma;\xi] 
\;,
\end{align}
where 
\begin{align}
\alpha=i(\chi -\nu-\mu)\;,\;\;
\beta= i(\chi -\nu+\mu)\;,\;\;
\gamma= 1-2i\nu
\;.
\end{align}
Plugging~\req{wform} into~\req{redefW} we obtain 
\begin{align}
\label{phifinal}
\phi=(-\xi)^{-i\nu}(1-\xi)^{i\chi }F_{21}[\alpha,\beta\,;\,\gamma;\xi]X^{\mathrm{W}}
\;,
\end{align}
which we apply in computation of the Sauter step potential effective action.

\section{Effect of $g\neq2$ on effective action for SS potential}
\label{periodicE}

We extend the derivation of SS effective action to $g\neq 2$, using the asymptotic states of the wave function~\req{phifinal}:
\begin{align}
\phi[z\to\infty]=&\;e^{ip_{z+}z}X^{\mathrm{W}} \;,
\nonumber \\
\phi[z\to-\infty]=&\;\Big(A e^{ip_{z-}z} + B e^{-ip_{z-}z}\Big)X^{\mathrm{W}} 
\;,
\end{align}
where the incident ($A$) and reflection ($B$) coefficients
\begin{align}
\label{incref}
A=&\;\frac{\Gamma[1-2i\nu]\Gamma[-2i\mu]}
{\Gamma[i(\chi -\nu-\mu)]\Gamma[1-i(\chi +\nu+\mu)]}\;,
\nonumber \\
B=&\;\frac{\Gamma[1-2i\nu]\Gamma[2i\mu]}
{\Gamma[i(\chi -\nu+\mu)]\Gamma[1-i(\chi +\nu-\mu)]}
\;.
\end{align}
Their ratio gives the Bogoliubov coefficient, but only the absolute value is needed for the imaginary part of the action: Following the notation of~\cite{Kim:2009pg}, 
\begin{align}
\label{Bogg}
\left|\frac AB\right|=\frac{\Gamma[i(\chi -\nu+\mu)]\Gamma[1-i(\chi +\nu-\mu)]}
{\Gamma[i(\chi -\nu-\mu)]\Gamma[1-i(\chi +\nu+\mu)]}
\;,
\end{align}
omitting the phase factor $\Gamma[-2i\mu]/\Gamma[2i\mu]$.

The spin-$1/2$ imaginary  part of action per unit cross sectional area is obtained by summing Bogoliubov coefficients: 
\begin{align}
\label{Lsum}
\mathrm{Im}[\mL^{1/2}_{\mathrm{SSg}}]=&\;
-\sum_{E,p_\perp}\sum_{s=\pm1}
\ln[\,|A/B|\,]
\nonumber \\
=&\;
-\int_D
\frac{d\omega d^2p_\perp}{(2\pi)^3}
\sum_{s=\pm 1} \ln[\,|A/B|\,]
\;,
\end{align}
where subscript \lq SSg\rq\ denotes the SS action with $g$-dependence, and $D$ denotes the domain of states capable of tunneling through the mass gap:
\begin{align}
\label{Dform}
\int_D d\omega d^2p_\perp =&\;
\int_{-(e\mE_0L/2-m)}^{e\mE_0L/2-m} \!\!\!\!\! d\omega
\int_0^{\sqrt{(|\omega|-e\mE_0 L/2)^2-m^2}} \!\! \!\!\! d^2p_\perp 
\;.
\end{align}

It suffices for $g\to2$ to set according to \req{chiform} $\chi \to s\lambda$ in \req{Bogg} to confirm that $|A/B|$ in \req{Lsum} matches the spin-$1/2$ expression in~\cite{Kim:2009pg}. Similarly, one can obtain spin-$0$ particle effective action taking $g\to0$ in $\chi $, \req{chiform}. This supports the notion that \req{chiform} is the correct analytical function of $g$. This result is directly dependent on our use of Klein-Gordon Pauli equation to describe spin-$1/2$ particles for $g\ne2$.

We simplify~\req{Lsum} by summing spin states and using $\chi$ relation \req{conj}. The resultant  gamma function can then be further simplified using the Euler reflection formula, {\it e.g.} 
\begin{align}
&\Gamma[i(\chi[s]-\nu+\mu)]\Gamma[1+i(\chi^*[-s]+\nu-\mu)]
\nonumber \\
&\;=\frac{-i\pi}{\sinh[\pi (\chi[s]-\nu+\mu)]}
\;.
\end{align}
This allows us to write the imaginary  part of action alone in terms of $\sinh$ functions
\begin{align}
\label{KleinIm}
\mathrm{Im}[\mL^{1/2}_{\mathrm{SSg}}]=&\;
\!-\!\!\int_D
\!\frac{d\omega d^2p_\perp}{(2\pi)^3}
 \\ \nonumber
 &\!\!\!\!\!\!\!\!\!\!\!\!\!\!\!\! \!\!\!\!\!\!\!\!\!\!\!\!\!\!\! \times 
\ln\! \Big[\Big|
\frac
{\sinh[\pi(\chi[|s|]-\nu-\mu)]\sinh[\pi(\chi[|s|]+\nu+\mu)]}
{\sinh[\pi(\chi[|s|]-\nu+\mu)]\sinh[\pi(\chi[|s|]+\nu-\mu)]}
\Big|\Big]
.
\end{align}  
In the limit $g\to 2$ using \req{chiform} with $\chi\to s\lambda$ \req{KleinIm} simplifies to Eq.\,(23) of~\cite{Kim:2009pg}, that is
\begin{align}
\label{KleinIm2}
\mathrm{Im}[\mL^{1/2}_{\mathrm{SS}}]=&\;
-\int_D
\frac{d\omega d^2p_\perp}{(2\pi)^3}
 \\ \nonumber
 &\!\!\!\!\!\!\!\!\!\!\!\!\!\!\!\! \times 
\ln\Big[\frac{\sinh[\pi(\lambda-\nu-\mu)]\sinh[\pi(\lambda+\nu+\mu)]}
{\sinh[\pi(\lambda-\nu+\mu)]\sinh[\pi(\lambda+\nu-\mu)]}\Big]
\;.
\end{align}

In order to amplify on the periodicity in $g$ of the imaginary part of the effective action we study the ratio\\[-0.8cm]
\begin{align}
\label{Rperiodic}
R=\frac{\mathrm{Im}[\mL^{1/2}_{\mathrm{SSg}}]}{\mathrm{Im}[\mL^{1/2}_{\mathrm{SS}}]}
\;,
\end{align}
of SSg~\req{KleinIm} with SS($g=2$)~\req{KleinIm2} actions normalizing the $g$-dependent action to the $g=2$ result. 

%
\begin{centering}
\begin{figure*}
\includegraphics[width=0.99\textwidth]{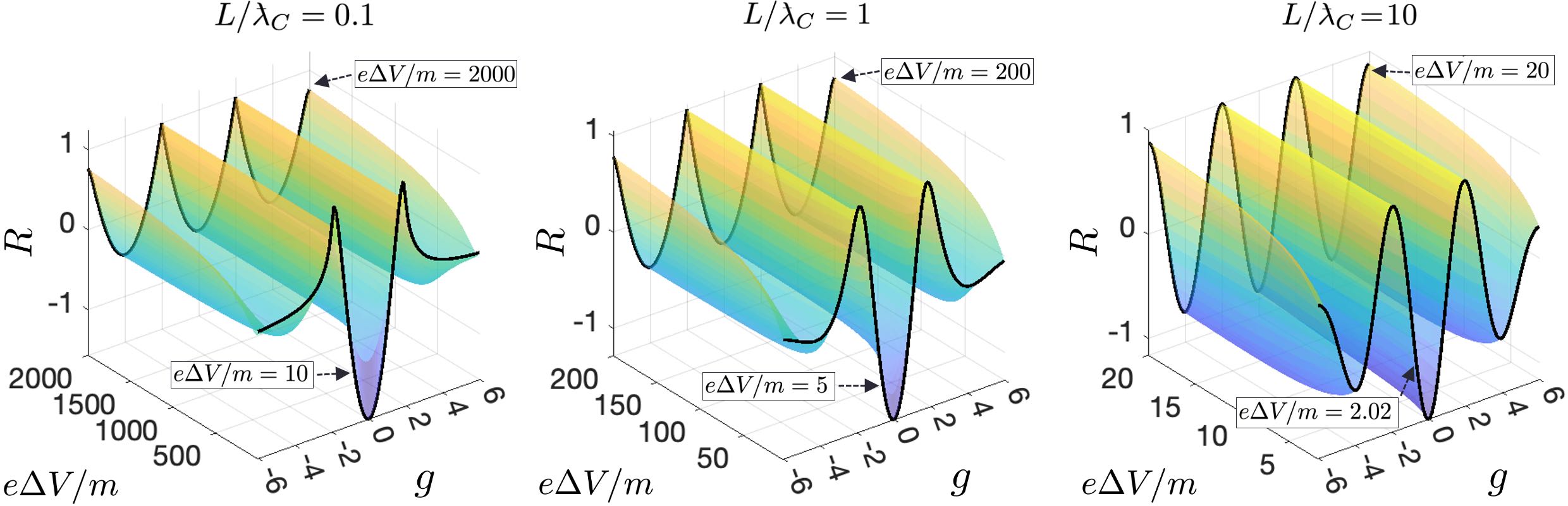}
\caption{\label{FigG} 
Plot of $R$ given by~\req{Rperiodic}: The imaginary part of the SSg effective action normalized to its $g=2$ value. We consider three different step widths $L$, for the given domains of $g$ and step height $e\Delta V$. Solid lines highlight the $g$-dependence at the boundaries of $e\Delta V$ domains.} 
\end{figure*}
\end{centering}
%
%

In~\rf{FigG} the ratio $R$ for the three different step widths $L/\lambdabar_\mathrm{C}=0.1,\; 1,\; 10$ ($\lambdabar_\mathrm{C}=\hbar/mc$) is shown as a function of $e\Delta V/m$ and $g$. While the range of $-6\le g\le 6$ displayed is fixed, the range of step heights $e\Delta V$ varies inversely with the scale of $L/\lambdabar_\mathrm{C}$. On the RHS ($L=10\lambdabar_\mathrm{C}$), the periodicity of imaginary part of effective action emerges at step height $e\Delta V= 20 m$, corresponding to a field strength at the center of the Sauter step of $\mE=2m^2/e$, twice the EHS critical field. As for more sharply peaked fields (middle and LHS in~\rf{FigG} with smaller $L$), the step height must be much greater in order for the imaginary part of SS action to exhibit periodicity in $g$. Note that at $g=\pm 2$ the periodic $g$-dependence is smooth at relatively weaker fields {\it i.e.} $2m^2/e$ (RHS), but develops cusps at much stronger fields, see {\it e.g.} the center plot at $e\Delta V=200m$ and $L=\lambdabar_\mathrm{C}$, corresponding to a peak electric field strength of $200m^2/e$, compare \req{ESauter}.

We find that periodicity in $g$ for the imaginary part of the action emerges in field configurations for which the SS effective action is compatible with the locally-constant EHS limit. This comparison takes into account the localized nature of the SS. Compatibility with EHS depends on both the potential step width and height. In general the compatible domain lies within a finite optimal width $L$ that depends on $e\Delta V$. The EHS limit is more accessible for larger $e\Delta V$, where a larger range of $L$ is permitted, Fig. 3 of~\cite{Evans:2021coy}. Consequently both subcritical fields $\mE\ll m^2/e$ with $L\gg\lambdabar_\mathrm{C}$, and stronger fields with $L<\lambdabar_\mathrm{C}$ can be compatible so long as $e\Delta V$ is large enough.

The periodic EHS result arises in \rf{FigG} for the three solid lines marked  $e\Delta V= 2000m$,  $e\Delta V= 200m$ and  $e\Delta V= 20m$ at their respective $L$ values. Periodicity for the imaginary part of the action surprisingly also arises for potential step configurations in which the SS action is incompatible with the EHS approximation. Specifically, the barely critical case $L=10\lambdabar_\mathrm{C}$ and $e\Delta V= 2.02m$ is closer to being periodic than the $L=\lambdabar_\mathrm{C}$ and $e\Delta V= 5m$ case, despite the former being the less compatible with EHS action (by orders of magnitude) than the latter configuration. 

\section{Periodic beta-function}
\label{betaCusp}

This quasi-constant EHS-gQED (EHSg) action allows for a straightforward evaluation of the beta-function using the strong field limit method summarized in~\cite{Dunne:2004nc}. We recall the magnetic dominated periodic in $g$ action from prior work~\cite{Rafelski:2012ui, Evans:2018kor} 
\begin{align}
\label{BEHSg}
\mL_\mathrm{EHSg}^{1/2}=&\,-\frac{m^4}{8\pi^2}\int_{m^2/\Lambda^2}^\infty\frac{dt}{t^3}e^{-t}
\\ \nonumber 
&\!\!\!\!\!\!\!\!\!\!\!\! \times 
\Big( \frac{t}{2z} \frac{\cosh[ g_kt/(4z)]}{\sinh[t/(2z)]} -1
-\frac{t^2}{12 z^2} \Big(\frac{3g_k^2}{8}-\frac12\Big)\Big)
\;,
\end{align}
where\\[-1cm] 
\begin{align} 
z=m^2/2e\mB
\;.
\end{align}
\req{BEHSg} is valid for all $g$ due to periodicity in the Weisskopf Landau level summation that implies a reset to the reduced domain  $|g|\le 2$:
\begin{align} 
\label{greset}
-2<g_k=g-4k <2\;,\quad k=0,\pm1,\pm2,\ldots 
\;,
\end{align}
ensuring a convergent proper time integral. As an example, the electron's $g=2+\alpha/\pi+\mathcal{O}(\alpha^2)$ resets to $g_{k=1}=g-4=-2+\alpha/\pi+\mathcal{O}(\alpha^2)$.
 
For the purpose of extracting the strong field EHSg limit of \req{BEHSg} we extend the $g=2$ work reviewed in~\cite{Dunne:2004nc}, to $|g_k|\le 2$. Using \textit{Mathematica} we obtain the generalization of the Hurwitz zeta function relation 
\begin{align}
\label{zetaform}
&\;
\frac{(s-1)^{-1}}{\Gamma[s-1]}\int_0^\infty \frac{dt}{t^{1-s}}e^{-2zt}
\\ \nonumber 
&\;\qquad \times 
\Big(\frac{\cosh[(g_k/2)t]}{\sinh[t]}-\frac1t-\frac t3\Big(\frac{3g_k^2}{8}-\frac12\Big)\Big) 
\\ \nonumber
&\ =\;
2^{-1-s}
\Big\{z^{-1-s}\Big(\frac{s}3\Big(\frac{3g_k^2}8-\frac12\Big) -\frac{4z^4}{s-1}\Big)
\\ \nonumber
&\; \qquad -2\Big(\zeta\Big[s,z-\frac{g_k-2}4\Big]+\zeta\Big[s,z+\frac{g_k+2}4\Big]\Big)\Big\}
\;.
\end{align}
Differentiating \req{zetaform} with respect to the variable $s$ as is done in~\cite{Dunne:2004nc}, and plugging the resulting integral equality into \req{BEHSg}, we obtain
\begin{align}
\label{3.22bHurwitz}
\mL_\mathrm{EHSg}^{1/2}=&\;
\frac{e^2\mB^2}{2\pi^2}\Big\{-\frac1{12}\ln[z]\Big(\frac{3g_k^2}8 -\frac12\Big) 
\\ \nonumber 
&\; 
\!\!\!\!\!\! + \frac12\Big( \zeta^\prime[-1,z+\frac{2-g_k}4]+\zeta^\prime[-1,z+\frac{2+g_k}4] \Big) 
\\ \nonumber 
&\; 
-\frac {3g_k^2/4-1}{24} +\frac{z^2}{4} -\frac{z^2}2\ln[z]
\Big\}
\;,
\end{align} 
where\\[-1cm]
\begin{align}
\zeta^\prime[-1,x]=\frac{\partial \zeta[s,x]}{\partial s}\Big|_{s=-1}
\;.
\end{align}
Setting $g_k=2$ we recover known result~\cite{Dunne:2004nc}.

This result \req{3.22bHurwitz} is particularly suitable for studying the very strong field limit.  In the strong field $e\mB\gg m^2 $ condition, the leading term in the first line of \req{3.22bHurwitz} dominates:
\begin{align}
\label{strongBZeta}
\mL_{\mathrm{EHSg}}[\mB]\to 
\frac{e^2 \mB^2}{24\pi}\Big(\frac{3g_k^2}8-\frac 12\Big)\ln[e\mB/m^2]
\;.
\end{align} 
The coefficient of the logarithmic term \req{strongBZeta} provides the beta-function $\beta$, from which we will extract the $g\neq2$-dependence. 

In order to establish the periodicity in $g$ of the $\beta$-function we need to use the periodicity we have proven for the imaginary part of action for the case of a pure electrical field. The imaginary part of the electric EHSg expression is obtained via  analytical continuation of the magnetic action \req{strongBZeta}. Substituting $\mB\to-i\mE$, we show that this representation of the effective action results in the periodic in $g$ imaginary part of the action we have presented in \rs{periodicE}:
\begin{align}
\label{strongEZeta}
\mL_{\mathrm{EHSg}}[\mE]\to 
-\frac{e^2 \mE^2}{24\pi}\Big(\frac{3g_k^2}8-\frac 12\Big)\ln[-ie\mE/m^2]
\;,
\end{align}
which reduces in the $g=2$ limit to
\begin{align}
\label{strongEZetag2}
\mL_{\mathrm{EHS}}[\mE]\to 
-\frac{e^2 \mE^2}{24\pi}\ln[-ie\mE/m^2]
\;.
\end{align}
With the ratio of~\req{strongEZeta} and~\req{strongEZetag2} we extract the $g$-dependent modification to the $\beta$-function:
\begin{align}
\label{RbetaF0}
\beta=\frac{e^3}{12\pi^2}R_\beta \;,\qquad
R_\beta=
\frac{\mathrm{Im}[\mL_{\mathrm{EHSg}}]_{e\mE/m^2\to\infty}}
{\mathrm{Im}[\mL_{\mathrm{EHS}}]_{e\mE/m^2\to\infty}}
\;,
\end{align}
where knowing the imaginary part of action suffices to obtain the $g$ contribution $R_\beta$:
\begin{align}
\label{RbetaF} 
R_\beta=&\;\frac{6}{\pi^2}\sum_{n=1}^\infty \frac{(-1)^n}{n^2}\cos[n\pi g/2]
\nonumber \\
=&\;
\begin{cases}
 3g^2/8-1/2\;, & |g|\le 2 \\
 3g_k^2/8-1/2\;, & |g|>2
 \end{cases} 
\;.
\end{align}
The $|g|\le 2$ case of~\req{RbetaF} reproduces the known solution obtained by Kruglov~\cite{Kruglov:2001dp}. At $g=2$ there is a cusp, and for the $|g|>2$ domain, $R_\beta$ reproduces the periodic beta-function from~\cite{Rafelski:2012ui}.

In~\rf{betaPeriodic} we validate $R_\beta$ in~\req{RbetaF} (solid) by plotting it alongside the localized field SSg result, $R$ from~\req{Rperiodic} (dashed, ratio of SSg to SS result). $R$ is plotted for the case $e\Delta V=200m$ and $L=\lambdabar_\mathrm{C}$, corresponding to maximum electric field strength $200m^2/e$. In this domain the Sauter step potential effective action is compatible with the strong field limit of the locally-constant EHS approximation, and there is excellent agreement with the periodic $g$-dependence predicted from $R_\beta$~\cite{Rafelski:2012ui}. The slight deviation between the two results at $g=\pm 6$ is due to the Sauter solution being a close but not exact match to the constant field solution. The trends we see in these results suggest that more precise in this parameter range numerical methods would allow for greater $e\Delta V$ values with an even closer agreement. 

%
\begin{figure}[h]
\begin{centering}
\includegraphics[width=0.8\columnwidth]{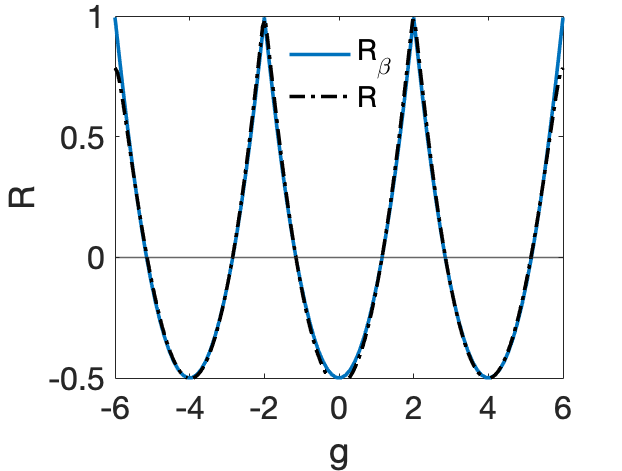}
\caption{\label{betaPeriodic} The $g$-dependent periodic beta-function~\req{RbetaF} (solid), alongside the SS potential effective action $R$ from~\req{Rperiodic} (dashed), with $L/\lambdabar_\mathrm{C}=1$ and $e\Delta V/m=200$.} 
\end{centering}
\end{figure}
%
%

For $g$ values far from 2  the gQED modification in~\rf{betaPeriodic} produces periodic sign flips in the beta-function. The sign change for $|g|\le 2$ in gQED result was noted previously~\cite{Kruglov:2001dp,Angeles-Martinez:2011wpn}: The beta-function is negative and thus signals infrared instability for $|g|<\sqrt{4/3}$. We confirm this result using an entirely different method, see~\rf{betaPeriodic}, and  demonstrate the $g$-periodicity: The same infrared instability arises for all $k$ in the intervals $-\sqrt{4/3}\pm4k<g<\sqrt{4/3}\pm4k$ according to periodic reset in~\req{greset}.

 As a verification of our approach, we check  that the periodic $R_\beta$ based on  $g$-reset effective action is consistent with the 2-loop QED corrections to the $\beta$-function. In context of QED diagram summation, the gQED result accounts exclusively for the internal photon vertex contributions to $\beta$, a subset of the irreducible diagrams that also consist of the self-energy corrections. By plugging the leading $\alpha$ correction $g=2+\alpha/\pi$ into $R_\beta$ in~\req{RbetaF}, we obtain the vertex contribution to the 2-loop $\beta$-function, that is a single internal photon vertex correction to the 1-loop vacuum polarization diagram. 

That the periodic reset is necessary for consistency can be seen by attempting to apply the electron's $g=2+\alpha/\pi$ (larger than 2) directly to $3g^2/8-1/2$, that is~\req{RbetaF} without reset, which would give $R_\beta=1+3\alpha/2\pi+\mathcal{O}(\alpha^2)$. Instead by accounting for periodic reset $g_k=g-4=-2+\alpha/\pi$, the leading $\alpha$ term flips sign in $R_\beta=1-3\alpha/2\pi+\mathcal{O}(\alpha^2)$. The latter gives the correct vertex contribution: negative and opposite in sign with respect to the self-energy term~\cite{Itzykson:1980rh}. This ensures that the gQED and self-energy contributions add to the full 2-loop result $R^{2-\mathrm{loop}}_\beta=1-3\alpha/4\pi+\mathcal{O}(\alpha^2)$, negative prior to insertion of mass counter terms.

\section{Summary and Outlook}
\label{sConcl}

We have presented modification of the imaginary part of gQED action for the electric Sauter step potential, see~\rf{FigG}, demonstrating that the effective action is convergent in the $|g|>2$ domain. This result extends the Schwinger proper time formulation which yields a convergent proper time integral only for $|g|\le 2$.   One of the key results of this work is that  we find the same periodicity in $g$ arises in the homogenous {\bf electric} field limit of the SS potential as in the {\bf magnetic} Weisskopf EHS summation study~\cite{Rafelski:2012ui}. Unlike the magnetic case, the periodicity in the electric-dominated action is not self-evident: The continuous spectra of Sauter solutions is not explicitly periodic until the Bogoliubov coefficient summation is performed. Our result also provides confirmation that the analytical extension proposed in Ref.~\cite{Rafelski:2012ui} to the case of arbitrary homogeneous and uniform electromagnetic fields applies.

We used this effective action periodicity to present the generalization of the Schwinger EHS effective action to $|g| > 2$. With this result we obtained the $g$-dependent modification of the beta-function, and verified it using the electric Sauter potential effective action, see~\rf{betaPeriodic}.  We found the same cusp at $g=2$ as in the magnetic case. Using an entirely different method in~\rs{betaCusp} when compared to prior art~\cite{Kruglov:2001dp,Angeles-Martinez:2011wpn}, we  also found the  domains of $g$ in which the beta-function is negative: These include, beyond the range $|g|<\sqrt{4/3}$ from Refs.~\cite{Kruglov:2001dp,Angeles-Martinez:2011wpn}, all periodic domains with $g$ reset by $\pm 4k$. These infrared unstable $g$-domains are indicating the appearance of a structured magnetic dominated charge confining quantum vacuum state, akin to the case of QCD, and will require a regularization procedure akin to that employed in the Savvidy action formulation~\cite{Savvidy:1977as}.

The $g$-dependent effective action offers an improved framework for exploring nonperturbative vacuum response. The imaginary  part of action result in~\rs{periodicE} helps us understand how to compute the real part of action. Based on the periodic beta function in~\rs{betaCusp}, we recognize how the $g$-dependence applies toward the renormalization subtraction required in the study of the real part of effective action. Based on Kim and Schubert's observation that the real part of action is uniquely determined by the structure of the imaginary part~\cite{Kim:2016xvg}, we expect the same periodicity to apply after renormalization to the finite real part. We hope to apply $g$-dependence to the real part of the SS potential effective action, implementing the $g=2$ result of Kim, Lee and Yoon~\cite{Kim:2009pg}.

The real part will allow us to extend the temperature representation of EHS action~\cite{Muller:1977mm, PauchyHwang:2009rz, Labun:2012jf} to $|g|>2$, and for further comparison to the Unruh temperature~\cite{Unruh:1976db}. This also paves the way to exploration of the interplay between gQED (irreducible) corrections and the higher order reducible loop corrections~\cite{Gies:2016yaa, Karbstein:2017gsb, Karbstein:2019wmj}. Here the real and imaginary parts of action mix in a nontrivial manner, see~\cite{Fedotov:2022ely} for a recent review.

In this work we have shown how $g$ affects the pure $\mE$ field action, which pertains to the scalar invariant $\mE^2-\mB^2$. To complete the understanding of effective action as a function of $g$, consideration of the case of nonvanishing pseudoscalar $\mE\cdot\mB$ is required.  Our result provides a framework for future study of $\mE\cdot\mB$ to all orders. We can utilize the approach seen in Euler and Heisenberg~\cite{Heisenberg:1935qt}, where the case of parallel electric and magnetic fields was studied. The pseudoscalar case is especially interesting because a cusp at $g=2$ at arbitrary field strengths was predicted~\cite{Rafelski:2012ui}, in contrast to the pure electric case where the cusp arises   in the asymptotic strong field limit governing the $\beta$-function.

An alternative method to the here employed second order KGP formulation is the first order DP equation. Comparing the KGP action to the DP results computed in the domain $|g|\le 2$~\cite{OConnell:1968spc, Dittrich:1977ee}, the weak field limits are compatible, but in stronger fields the energy eigenvalues and thus the effective actions can differ significantly~\cite{Steinmetz:2018ryf}. It is important to note that numerous DP results were obtained and discussed prior to this work -- these include in particular the case of neutral particles, where DP action was evaluated in quasi-constant fields relevant to magnetars~\cite{Ferrer:2019xlr} and in a magnetic Sauter step~\cite{Adorno:2021xvj}. Our work showing periodicity in $g$ suggests a need to revisit these results with the KGP method.

The imaginary part of effective action here presented for a Sauter step potential describes production of particle pairs at the surfaces of extreme field stelar objects such as magnetars~\cite{Ruffini:2009hg} and may help improve models of strong field energy conversion into particle pairs considered a potential source of gamma ray burst energy (GRB)~\cite{Moradi:2021hus}. We further recall that in magnetar fields a small change in $g$ at perturbative QED $\alpha/\pi$ scale suffices to suppress EHS pair production by factor $10^{-3}$~\cite{Evans:2018kor}. Our inhomogeneous field results further reduce the pair production rate opening to review the pair production mechanisms of GRB formation.

The presence of free particles at large distances is a natural consequence of our study of  the finite extent in space of the SS. An incident particle's charge and magnetic dipole current are conserved as it travels across the SS space domain; magnetic moment $\mu$ is independent of the external  SS. We recall $g\propto m\mu$, and it is known that mass is modified by externally applied fields~\cite{Ritus:1972ky,Evans:2019zyk}. This implies that $g$ and $m$ respond equally in the presence of an external field. A suitable probe for this field-dependence is the particle production rate, which is sensitive to both these parameters~\cite{Evans:2018kor}.

Our improved theoretical formulation of gQED action should allow for better understanding of how anomalous moment affects the vacuum response. The periodic in $g$ result and cusp structure could lead to a renewed search for nonperturbative QED vacuum effects.


\end{document}